\documentclass[conference,a4paper]{IEEEtran}
\usepackage{graphicx,color}
\usepackage{amssymb}
\usepackage{tabularx}
\usepackage{subfigure}

\newtheorem{theorem}{Theorem}

\newtheorem{lemma}{Lemma}

\newtheorem{example}{Example}

\begin{document}

\sloppy

\title{Using Feedback for Secrecy over Graphs} 
\author{
  \IEEEauthorblockN{
    Shaunak Mishra\IEEEauthorrefmark{1}, Christina Fragouli\IEEEauthorrefmark{2}, Vinod Prabhakaran\IEEEauthorrefmark{3} and Suhas Diggavi\IEEEauthorrefmark{1}}
  \IEEEauthorblockA{
    \IEEEauthorrefmark{1}%Electrical Engineering Department\\
    University of California, Los Angeles, USA}%\\ 
    %Email: shaunakmishra@ucla.edu}
  \IEEEauthorblockA{
    \IEEEauthorrefmark{2}%School of Technology and Computer Science\\
  Tata Institute of Fundamental Research, India\\
    % Email: vinodmp@tifr.res.in
}
  \IEEEauthorblockA{
    \IEEEauthorrefmark{3} %School of Computer and Communication Sciences\\
Ecole Polytechnique Federale de Lausanne, Switzerland\\
 %   Email: christina.fragouli@epfl.ch
}
 }

%% Create the title:
\maketitle

\begin{abstract} We study the problem of secure message multicasting over graphs in the presence of a passive (node) adversary who tries to eavesdrop in the network. We show that use of feedback, facilitated through the existence of cycles or undirected edges, enables higher rates than possible in directed acyclic graphs of the same mincut. We demonstrate this using code constructions for canonical combination networks (CCNs). We also provide general outer bounds as well as schemes for node adversaries over CCNs.
\end{abstract}

\section{Introduction}
%=============================
Consider a source that would like to securely multicast a message 
to a set of receivers in the presence of passive adversaries.
It is well known that over wireless networks, if public feedback 
is available,  we can support  higher secrecy rates than if it is not \cite{maurer}.
We explore in this paper whether the same could be true over wired networks that are modeled as graphs.

While security against eavesdropping has been extensively examined (in a number of interesting works) 
in the network coding literature, the potential utility
of feedback as such has not, as far as we know. 
Seminal works such as \cite{secure_network_coding,koetter} have looked both at information theoretical bounds
as well as code constructions for the case of edge adversaries;  works have also started examining
the case of node adversaries \cite{netcod_12_ghid,kosut}. In all cases however the underlying network is modeled as
a directed acyclic graph.

Yet feedback is readily available in wired networks, and could potentially help in secrecy. 
Many times connections between sources and receivers 
are undirected or bi-directional; even over directed graphs, we may have cycles, that offer a form of feedback between
network nodes. The existence of such cycles could be put to good use to create for instance common randomness
between intermediate network nodes, that a secrecy protocol could leverage to achieve higher rates.

We here provide a number of examples to establish that this is indeed the case.
We mainly consider node adversaries, that tap a specific network node and
intercept all incoming messages, but also discuss  edges adversaries.
%For node adversaries no general results are known, even without feedback.
We  focus on a special class of (minimal) combination networks, that is often used in the network coding literature,
and the simplest possible case, of a single node adversary.
We derive outer bounds as well as achievability schemes for the cases where feedback
is (and is not) available. We design schemes that employ feedback, which can offer rates higher
than outer bounds in the case where feedback is not available. These results point to the potential of using such feedback for network secrecy; a topic of ongoing investigation.

The paper is organized as follows. Section~\ref{sec:setup} introduces our notation and basic notions; Section~\ref{sec:SMT_rate_results} examines feedback over very simple abstracted examples. Section~\ref{sec:without_feedback} deals with inner and outer bounds for directed acyclic graphs, mainly developed for comparison purposes in this paper. Finally, Section~\ref{sec:with_feedback} shows the benefits of feedback in undirected and bidirected graphs.

\section{Notation and setup} \label{sec:setup}
%====================================================
We model a wired network as  a graph $ \mathcal{G}= (\mathcal{V},\mathcal{E})$ with unit capacity edges. A single source node has a message 
 $\mathcal{W}$  to send to a set of receivers $\mathcal{R} \subset \mathcal{V}$. 
 We are interested in deriving outer bounds, as well as building
 $N$-round secure protocols with the following constraints:\\
%\begin{itemize}
%\item 
$\bullet$ \textit{Round} : Each edge can be used at most once in a round.\\
%$\bullet$  \textit{Linear} : We allow intermediate nodes to do linear operations over a finite field  $\mathbb{F}$.\\--------------------------DOUBT------------------------
%\item 
$\bullet$ \textit{Decodability} : All receivers \textit{perfectly} decode message $\mathcal{W}$ with zero error probability.\\
%\item 
$\bullet$ \textit{Secrecy} : $ H(\mathcal{W}|\mathcal{V}_{\mathcal{A}}) =H(\mathcal{W})$
where $\mathcal{V}_{\mathcal{A}}$ denotes the ``view'' of an adversary $\mathcal{A}$, i.e, the information available to tapped edges or nodes during the protocol.\\ 
%\end{itemize}
We say that such a protocol achieves\\
$\bullet$ \textit{Secrecy rate} : $\frac{H(\mathcal{W})}{N}$.\\
We distinguish between two types of passive adversaries :\\
$\bullet$ \textit{k-edge} adversary : the adversary has access to an arbitrary set of $k$ edges.\\
 $\bullet$ \textit{k-node} adversary : the adversary has access to an arbitrary set of $k$ nodes. In this paper, we mainly focus on a $1$-node adversary.\\
We allow intermediate nodes to do operations over a finite field  $\mathbb{F}$. We also assume that the  network nodes share no prior common randomness and no side secure communication channel, they can only communicate through the network graph that is subject to eavesdropping.

\begin{figure}[!th]
\begin{center}
\includegraphics[width=2.67in,height=2.6in]{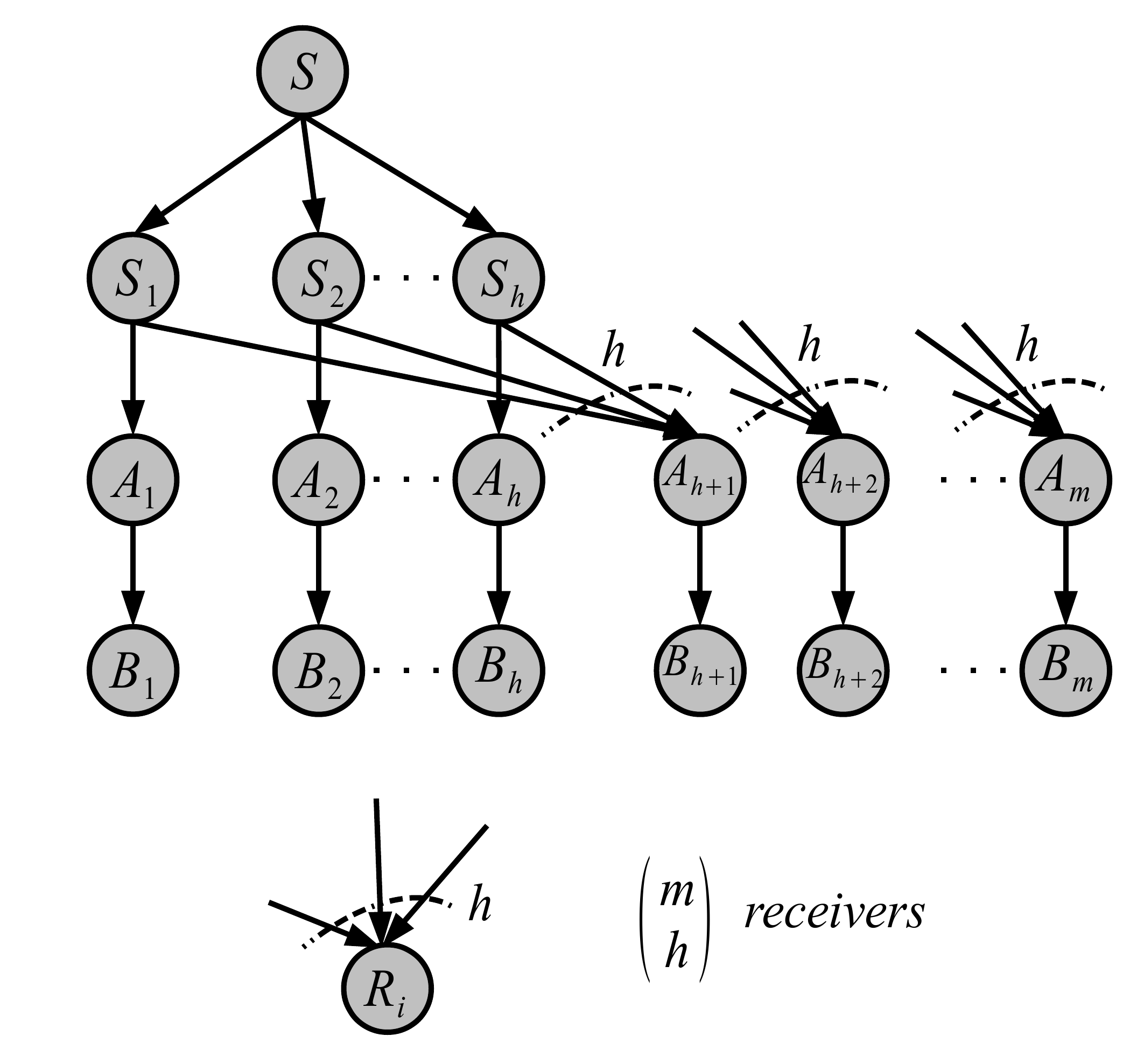}
\caption{Directed $(m,h)$-CCN.}
\label{fig:CCN_directed}
\end{center}
\end{figure}

\paragraph*{Canonical combination networks (CCNs)}Our results in this paper are over CCNs, that essentially are minimal\footnote{Removing any edge reduces the mincut for at least one receiver.} combination networks, see \cite{NCmonograph,netcod_12_ghid}. Figure~\ref{fig:CCN_directed} shows a {\em directed} $(m,h)$-CCN with $m\geq h$, where $m$ is the number of coding points and $h$ the mincut to the receivers. It has a source $S$, $h$ trivial coding nodes $A_1,A_2,\ldots,A_h$ (with indegree one), $m-h$ non-trivial coding nodes $A_{h+1},A_{h+2},\ldots,A_{m}$ (with indegree $h$) and $m \choose h$ receivers. Each receiver is connected to $h$ nodes from the set $\{B_1, B_2,\ldots,B_m\}$. An {\em undirected} $(m,h)$-CCN can be obtained by replacing all the directed edges in the directed $(m,h)$-CCN by undirected edges. We will also consider a  {\em bidirected} $(m,h)$-CCN which we create by 
adding an edge (backward edge) in the reverse direction for every edge (forward edge) in the directed $(m,h)$-CCN. Note that the directed, undirected and bidirected networks all have mincut $h$ towards each receiver.

\begin{figure}[!th]
\begin{center}
\includegraphics[width=3.3in,height=1.52in]{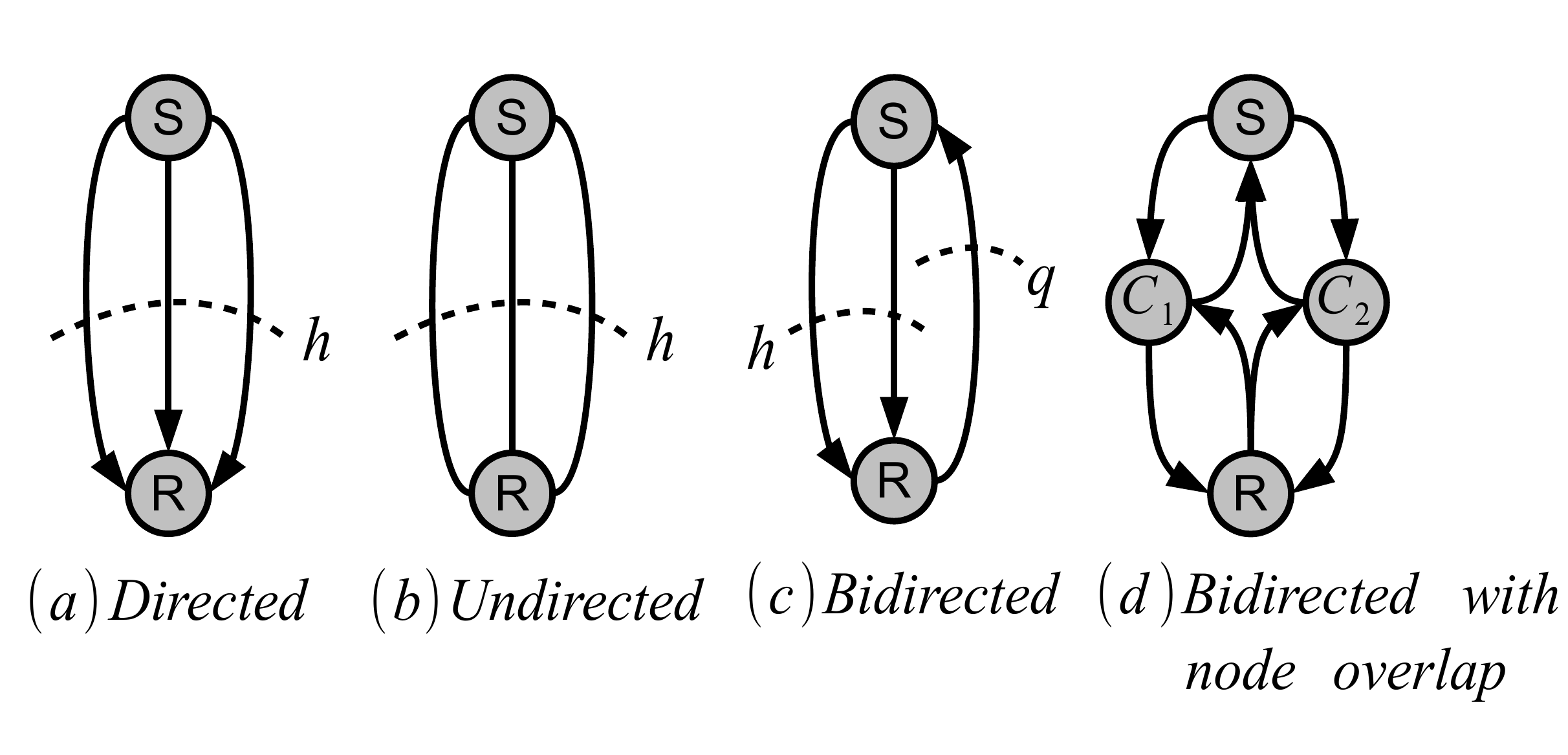}
\caption{Abstracted examples. The cut values are shown beside dotted lines for (a), (b) and (c).} %are accompanied by the cardinality of edges in the cut.}
\label{fig:SMT_cases}
\end{center}
\end{figure}

\section{Illustrating examples} \label{sec:SMT_rate_results}
We consider the simple networks depicted in Figure~\ref{fig:SMT_cases},
where a single source $S$ wants to securely send information to a receiver $R$; we think of the edges in these networks
as abstracting edge-disjoint (and in the last case node-disjoint) paths in larger networks.
Our goal is to build  basic intuition on when feedback could be useful.
For the first three examples we assume a $k$-edge adversary, while for the last a $1$-node adversary.

\paragraph*{Example in Figure~\ref{fig:SMT_cases}(a)} the outer bound on the secrecy rate is $h-k$ and is achievable. This follows trivially from \cite{secure_network_coding}.

\paragraph*{Example in Figure~\ref{fig:SMT_cases}(b)} for this undirected graph the  outer bound remains $h-k$ (proved below), and its achievability follows from \cite{secure_network_coding}. We can show the outer bound as follows.
We apply the  ``crypto-inequality'' in \cite{vazirani} which states that
\begin{eqnarray}
I(\mathcal{W},\; \mathcal{K}_S; \mathcal{K}_R \;| \; U^N) = 0 \label{crypto_inequality}
\end{eqnarray}
where $\mathcal{W}$ is the source message, $\mathcal{K}_S$  is the private randomness of the source, $\mathcal{K}_R$ is the private randomness of the receiver and $U^N$ denotes the values exchanged between the source and receiver during any $N$-round protocol. Simply put, (\ref{crypto_inequality}) implies that $(\mathcal{W},\; \mathcal{K}_S)$ and $\mathcal{K}_R$, which were independent to begin with, remain independent even after conditioning on all values exchanged during the protocol. For the graph in this example, 
\begin{eqnarray} 
H(\mathcal{W}) \stackrel{(a)} \leq I(\mathcal{W}; U^N_{1:h} \mathcal{K}_R) \stackrel{(b)}= I(\mathcal{W}; U^N_{1:h}) \stackrel{(c)} \leq  N(h-k) \label{eq:SMT_undirected}
\end{eqnarray}
where $U^N_{1:h}$ denotes\footnote{We use the notation $U^N_{1:h}$  for $U^N_1, U^N_2,\ldots,U^N_h$.} values exchanged between the source and receiver during the $N$-round protocol, (a) follows from the decodability constraint, (b) follows from (\ref{crypto_inequality}) and (c) follows from the secrecy constraint. %This can be simply be achieved using secure network coding \cite{secure_network_coding}.

\paragraph*{Example in Figure~\ref{fig:SMT_cases}(c)} for this cyclic graph, if we have $h$ forward (from $S$ to $R$) edges and $q$ backward edges (from $R$ to $S$),
the outer bound becomes $\min\{h,h+q-k\}$ where the bound $h+q-k$ follows easily with similar steps as in (\ref{eq:SMT_undirected}). The outer bound is achievable as follows. To achieve it, when $k\leq q$, we can send $q$ random packets (keys) from $R$ to $S$ using the backward edges, say $r_1,\; \ldots,\;r_q$. The source creates $h$ linear combinations of these $r$-packets (using for instance an MDS code), say $s_1,\; \ldots,\;s_h$ such that %\begin{itemize}
%\item 
the $r$-packets and the $s$-packets are in general position (any selection of $q$ of these packets are linearly independent). 
%\item In addition, if any $l$ $r$-packets ($l\leq q$) are set to zero in the $h$ linear combinations forming $s$-packets, any selection of $q-l$ of these linear combinations are still linearly independent.
%\end{itemize}
 The source uses the $s$-packets as one-time pads for the forward edges. With this construction, an adversary observing any $k$ edges will not be able to retrieve information (proved below) and thus secrecy rate $h$ is achievable. For proof of secrecy, consider a $k$-edge adversary who taps $l$ backward edges and $k-l$ forward edges. From $l$ backward edges it infers $l$ $r$-packets (say $r_1,\ldots,r_l$). On the forward edges, the adversary observes $\mathcal{V_A}$ (after accounting for inferred packets $r_1,\ldots,r_l$) as shown below,
%=============insert extended version data================================================
\begin{eqnarray}
\mathcal{V_A} =  \left( \matrix{ b_1 \cr b_2 \cr .\cr. \cr b_{k-l} }\right) + \mathbf{A} \left( \matrix{ r_{l+1} \cr r_{l+2} \cr .\cr. \cr r_{q} }\right)
%B = \left( \matrix{ b_1 \cr b_2 \cr .\cr. \cr b_ }\right) \quad V =  \left( \matrix{ b_1 \cr b_2 \cr .\cr. \cr b_k }\right)
\end{eqnarray}
where $b_1,\ldots,b_{k-l}$ are information symbols (on $k-l$ forward edges) and matrix $\mathbf{A}$ is full rank (by construction of $s$-packets).
Now, $H(b_1,\ldots,b_{k-l} | \mathcal{V_A}) = H(b_1,\ldots,b_{k-l})$ since $r$-packets are uniformly distributed, $\mathbf{A}$ is full rank and $k\leq q$.
%==========================================================
We can easily extend this scheme in the case where  $k > q$, by combining the previous scheme with the scheme in \cite{secure_network_coding}:
use again the backward edges to convey random packets to the source, have the source itself generate $k-q$ random packets, and combine these to create one-time pads to encode 
$h-(k-q)$ information messages to send to the receiver using secure network coding (for more details see \cite{secure_network_coding}).

\paragraph*{Example in Figure~\ref{fig:SMT_cases}(d)} for this bidirectional graph with two intermediate nodes $C_1$ and $C_2$, we can achieve secrecy rate $2$ even if there is a $1$-node passive adversary. To do so, the receiver $R$ can send keys $k_1$ and $k_2$ to the source $S$. Since $C_1$ and $C_2$ each observe only one of the keys, $S$ can use $k_1 +k_2$ as a one-time pad for both the forward paths. If we now have a network with $h$ overlapping forward and backward paths through intermediate nodes $C_1,C_2,\ldots,C_h$ (each forward path has a node-overlap with only one backward path), secrecy rate $h$ is again achievable against a $1$-node adversary using the same approach (this will be a technique we will use for $(m,h)$-bidirected CCN considered in a later section).

\paragraph*{\bf Intuition} These simple examples give a very intuitive message:
if we can use feedback (edges in backward directions, in cyclic graphs) without affecting the mincut,
then this can help to achieve higher\footnote{In addition to these examples, results for unicast in \cite{kamal_jain,desmedt_revisited} also show feedback leads to a relaxation in connectivity requirements for secure message transmission.} secrecy rates. Essentially, we can use the feedback to create common randomness. In Section~\ref{sec:with_feedback}, we show that this is indeed the case in more complex networks with multiple receivers as well, where however more elaborate schemes will be needed.

\section{Directed CCN with 1-node adversary} \label{sec:without_feedback}
%============================================================================
In this section, we derive outer bounds and achievability schemes for directed CCNs where a single node is compromised and acts as a passive adversary.
This is a case without feedback, which we mainly develop for comparison purposes, but we believe that these results are of independent interest.
We give first an outer bound and then achievability schemes that match the outer bound in some cases.

\begin{theorem} \label{thm:outer_bound_directed}
Consider a directed $(m,h)$-CCN with\footnote{We consider the case of directed CCNs with $m>h$  as in the trivial case of $m=h$ the outer bound against a $1$-node adversary is $h-1$ and is achievable using secure network coding \cite{secure_network_coding}.}
$m\geq h+1$. An outer bound on the secrecy rate against a $1$-node adversary is 
\begin{eqnarray}
\frac{(h-1)^2}{h}
\end{eqnarray} 
\end{theorem}
\begin{IEEEproof}
Consider a directed $(m,h)$-CCN with $m=h+1$. The high level idea of the proof will be to derive ``top'' and ``lower'' layer constraints for this layered network and then combine the two using Markovity relationships.
\par For an $N$-round protocol and $1\leq i \leq h$, let $Z_i^N$, $Y_i^N$ and $L_i^N$ denote the values sent on edge $S \rightarrow S_i$, $S_i \rightarrow A_{h+1}$ and $S_i \rightarrow A_{i}$ respectively. For $1\leq i \leq m$, let $T_i^N$ denote the values sent on edge $A_i \rightarrow B_{i}$. Figure~\ref{fig:butterfly_proof} illustrates the use of notation for a directed (3,2)-CCN.
\begin{figure}[!ht]
\begin{center}
\includegraphics[width=2.1in,height=2.4in]{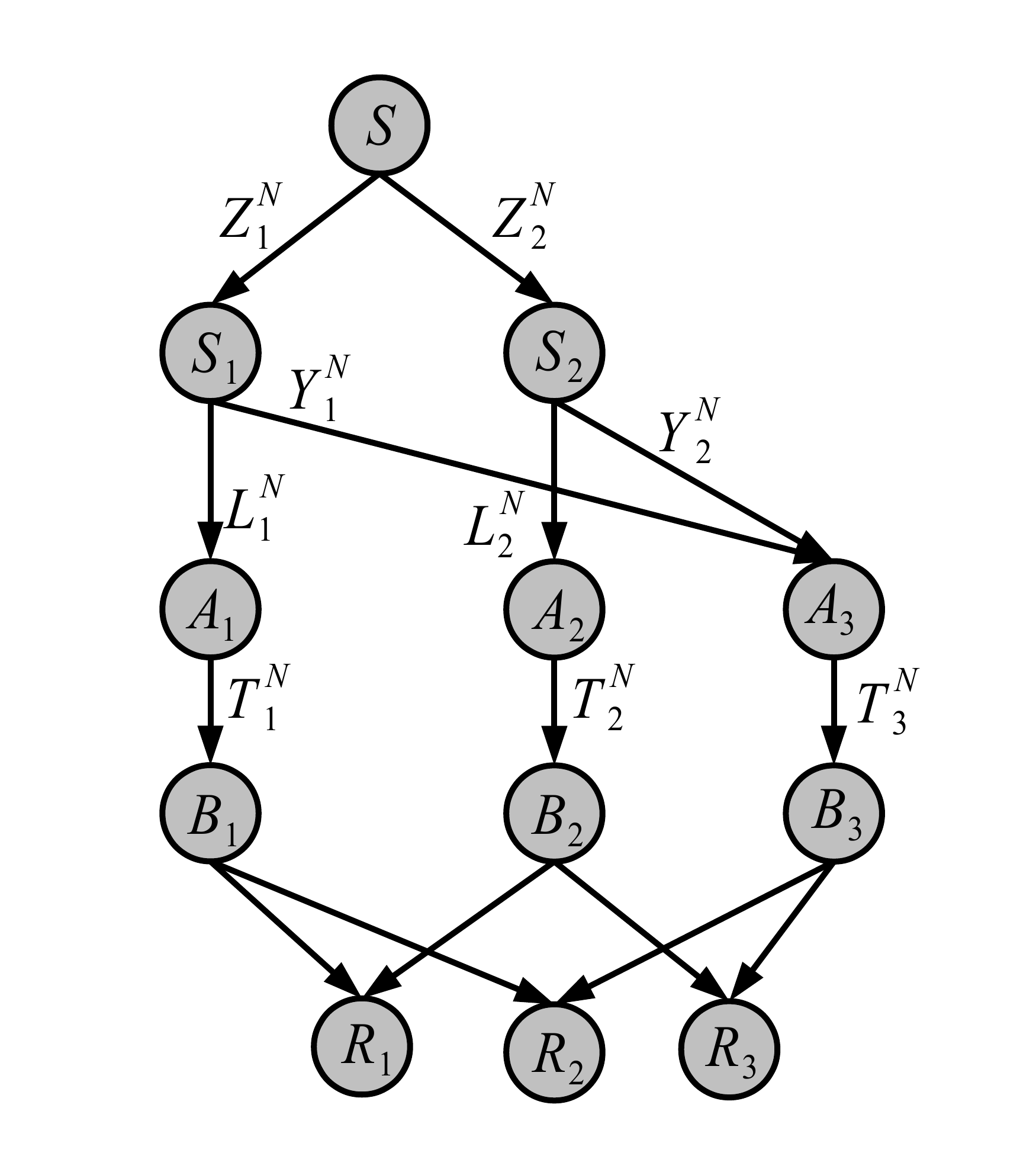}
\caption{A directed $(3,2)$-CCN illustrating notation used for proof of Theorem~\ref{thm:outer_bound_directed}.}
\label{fig:butterfly_proof}
\end{center}
\end{figure}
%The proof technique will be to derive ``top'' and ``lower'' layer constraints and combine the two using Markovity relationships.\\
\\\textit{Top layer constraints :}
\begin{eqnarray}
&& (h-1)N - H(\mathcal{W}) \nonumber \\ &\geq& I(\mathcal{W}Z_1^N; Z_{2:h}^N)- H(\mathcal{W})\nonumber\\
%&=&I(Z_1^N; Z_{2:h}^N) + \nonumber\\ &&  I(\mathcal{M}; Z_{2:h}^N |  Z_1^N ) \nonumber\\
%&=&I(Z_1^N; Z_{2:h}^N ) + H(\mathcal{M}| Z_1^N ) - \nonumber \\ && H(\mathcal{M} | Z_{1:h}^N) \nonumber\\
&\stackrel{(a)}=&I(Z_1^N; Z_{2:h}^N) + H(\mathcal{W}| Z_1^N )  - H(\mathcal{W}) \stackrel{(b)}=  I(Z_1^N; Z_{2:h}^N) \nonumber
\end{eqnarray}
where (a) and (b) follow from decodability %\footnote{This follows from Fano's inequality. We ignore the negligible error probability $\epsilon$ for notational convenience.}
and secrecy constraints. In general, for any $i\in\{1,2,...h\}$ we have the following result,
\begin{eqnarray} \label{eq:top_layer}
 I(Z_i^N; \{Z_j^N\}_{j\neq i}) &\leq (h-1)N -H(\mathcal{W})
\end{eqnarray}
where $\{Z_j^N\}_{j\neq i}$ denotes the set $\{Z_1^N,Z_2^N...Z_h^N\} - \{Z_i^N\}$.
\textit{Markovity :}
\begin{eqnarray} \label{eq:markov}
I( L_1^N  Y_1^N  ; L_{2:h}^N Y_{2:h}^N)&\leq&I(Y_1^N L_1^N Z_{1}^N;Y_{2:h}^N L_{2:h}^N Z_{2:h}^N) \nonumber\\
%&=&I(Z_1^N;Y_{2:h}^N L_{2:h}^N Z_{2:h}^N) + \nonumber\\ &&I(Y_1^N L_1^N ;Y_{2:h}^N L_{2:h}^N Z_{2:h}^N|Z_1^N) \nonumber \\
%&\stackrel{(a)}=&I(Z_1^N;Y_{2:h}^N L_{2:h}^N Z_{2:h}^N) \nonumber\\
%&=&I(Z_1^N;Z_{2:h}^N)  + \nonumber\\ && I(Z_1^N; Y_{2:h}^N L_{2:h}^N |Z_{2:h}^N)  \nonumber\\
&\stackrel{(a)}  = & I(Z_1^N;Z_{2:h}^N) 
\end{eqnarray}
where (a) follows from Markov chains $Y_{2:h}^N L_{2:h}^N Z_{2:h}^N \rightarrow Z_1^N \rightarrow Y_1^N L_1^N $ and $Y_{1}^N L_{1}^N Z_{1}^N \rightarrow Z_{2:h}^N \rightarrow Y_{2:h}^N L_{2:h}^N $. Similarly, $  I(L_i^N Y_i^N; \{L_j^N Y_j^N\}_{j\neq i}) \leq I(Z_i^N ; \{Z_j^N \}_{j\neq i})$.\\
\textit{Lower layer constraints :}
\begin{eqnarray}
&& H(\mathcal{W}) \nonumber \\
&=& I(\mathcal{W}; L_{1:h-1}^N Y_{1:h}^N)%\nonumber\\ 
\stackrel{(a)}= I(\mathcal{W}; L_{1:h-1}^N  | Y_{1:h}^N) \nonumber\\
&\leq& \sum_{i=1}^{h-1} H (L_i^N  | L^N_{1:i-1} Y_{1:h}^N) -  H( L_i^N  | L^N_{1:i-1} Y_{1:h}^N  \mathcal{W} L^N_{i+1:h} ) \nonumber\\
%&\stackrel{(b)}=& \sum_{i=1}^{h-1} ( H (L_i^N  | L^N_{1:i-1} Y_{1:h}^N) -  \nonumber\\ && H( L_i^N  | L^N_{1:i-1} Y_{1:h}^N  L^N_{i+1:h} ) )\nonumber\\
&\stackrel{(b)}=&\sum_{i=1}^{h-1}  I (L_i^N ;L^N_{i+1:h}  | L^N_{1:i-1} Y_{1:h}^N ) \nonumber\\
&\leq& \sum_{i=1}^{h-1}  I(L_i^N Y_i^N; \{L_j^N Y_j^N\}_{j\neq i})  \nonumber\\
&\stackrel{(c)}\leq &\sum_{i=1}^{h-1}  I(Z_i^N ; \{Z_j^N \}_{j\neq i}) \stackrel{(d)}\leq (h-1) ((h-1)N - H(\mathcal{W})) \nonumber
\end{eqnarray}
where (a) follows from secrecy constraint at node $A_{h+1}$, (b) follows from decodability constraint for the receiver not connected to node $B_i$, (c) follows from Markov chains as shown in (\ref{eq:markov}) and (d) follows from the top layer constraints (\ref{eq:top_layer}).
This completes the proof for $m=h+1$. The same outer bound holds for any directed $(m,h)$-CCN with $m>h$ due to the presence of receivers used in the proof for $m=h+1$.
\end{IEEEproof}
%Note that the above outer bound proof also holds when we allow a negligible probability of error in decoding the message.

\begin{lemma}\label{lemma:inner_tight_directed}
 \label{sec:inner_bound_directed}
Consider a directed $(m,h)$-CCN with a $1$-node adversary. There exist achievable schemes that 
are tight for the cases $(m , h=2)$, $(m=h+1,h)$ and $(m\leq 6, h=3)$.
\end{lemma}
\begin{IEEEproof} See Appendix.
\end{IEEEproof}

\par
\begin{figure}[!th]
\begin{center}
\includegraphics[width=1.815in,height=1.55in]{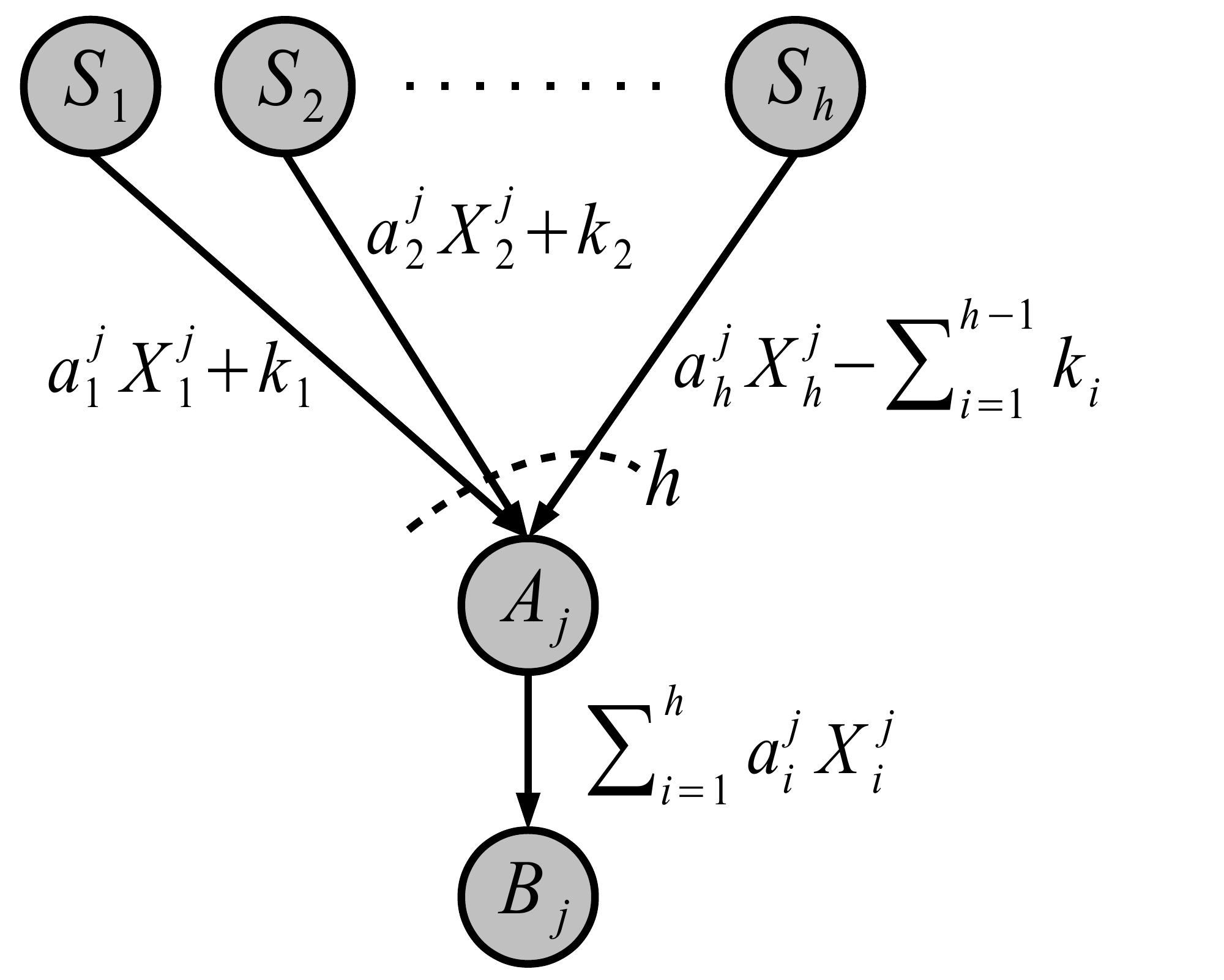}
\caption{Second round of KSC scheme.} %are accompanied by the cardinality of edges in the cut.}
\label{fig:KSC_round_2}
\end{center}
\end{figure}

 For a directed $(m,2)$-CCN, there exists an alternative optimal scheme (parts of which we will use in our feedback schemes in Section~\ref{sec:with_feedback}). It achieves secrecy rate $\frac{h-1}{2}$ in a directed $(m,h)$-CCN with $1$-node adversary. Described below, this scheme uses additional keys which do not reach the receivers and are \textit{cancelled} at intermediate nodes\footnote{This scheme is similar to achievability schemes in \cite{netcod_12_ghid}. In addition to \cite{netcod_12_ghid}, the approach of cancelling keys at intermediate nodes is also shown in \cite{tracey_ho}.}.
\paragraph*{Key set cancellation (KSC) scheme}
This is a $2$-round scheme.
\begin{enumerate}
\item In the first round, the source $S$ sends keys $k_1,k_2,\ldots,k_{h-1}$ to nodes $S_1,S_2,\ldots,S_{h-1}$ respectively and $k_h=-\sum_{i=1}^{h-1}k_i$ to $S_h$.
\item In the second round, a secure network code \cite{secure_network_coding} for $1$-edge adversary is used with a slight modification (described below) using keys from the first round.
This delivers $h-1$ symbols to all receivers in the second round and achieves secrecy rate $\frac{h-1}{2}$ over $2$ rounds.
\end{enumerate}
The modification mentioned in the second round is as follows.
Consider a secure network code \cite{secure_network_coding} for a directed $(m,h)$-CCN with $1$-edge adversary.
For this specific code, $\forall j>h$ let $X^j_i$ and $\sum_{i=1}^{h}a^j_i X^j_i$ be the values sent on edges $S_i \rightarrow A_j$ and  $A_j \rightarrow B_j$ respectively.
In the second round of KSC scheme, we use this code with the modification that $\forall j>h$, $S_i$ sends $a^j_iX^j_i + k_i$ to $A_j$ (non-trivial coding node) instead of $X^j_i$.
Node $A_j$ sums up all the values received from $S_1,S_2,\ldots,S_h$ (shown in Figure~\ref{fig:KSC_round_2}) and sends $\sum_{i=1}^{h}a^j_i X^j_i + \sum_{i=1}^{h}k_i = \sum_{i=1}^{h}a^j_i X^j_i $ to $B_j$.
Hence the \textit{key set} $\{k_1,k_2,\ldots,k_{h}\}$ accumulated in the first round is cancelled at all non-trivial coding nodes.
The key set ensures secrecy at every non-trivial coding node and the underlying secure network code delivers $h-1$ symbols to all receivers.

%%%%%%%%%%%%%%%   DUMMY TABLE %%%%%%%%%%%%%%%%%%%%%%%%%%%%%%%%%%%%%%%%%%%%%%%%%

%%%%%%%%%%%%%%%%%%%%%%%%%%%%%%%%%%%%%%%%%%%%%%%%%%%%%%%%%%%%%%%%%
\section{Undirected and bidirected CCN} \label{sec:with_feedback}

%  summarizes our results for a 1-node adversary.
%(** indicates a result that we were able to prove for a subset of the graphs).
In this section, we show that use of feedback can improve secrecy rates for undirected and bidirected graphs.
We prove in particular the results for undirected and bidirected CCNs
that are summarized in Table~\ref{table:summary}.
%%%%%%%%%%%%%%%%%%%%%%%%%%%%%%%%%%%%%%%%%%%%%%%%%%%%%%%
 \begin{table}[t]
\caption{Summary of results for 1-node adversary} 
\centering % title of Table
\begin{minipage}[c]{0.8 \linewidth}
\centering
\begin{tabular}{c  c  c} % centered columns (4 columns)
\hline     \\%inserts double horizontal lines
$(m,h)$-CCN &   Multicast & Multicast\\ %
 type        &    inner & outer \\
             &  bound & bound \\[0.5ex]
Directed\footnote{The multicast inner bound shown for directed CCN is for a limited number of cases described in Lemma~\ref{lemma:inner_tight_directed}.}  & ${\frac{(h-1)^2}{h}}$ &$\frac{(h-1)^2}{h}$  \\ 
Undirected & $(h-1)\frac{(m-h+1)}{(m-h+2)}$ &  $h-1$\\ 
Bidirected  & $h-1$ & $h$\\ [1ex]      % [1ex] adds vertical space
\hline %inserts single line
\end{tabular}
\end{minipage}
\label{table:summary} % is used to refer this table in the text
\end{table}
\subsection{Undirected CCN}

An undirected CCN allows the usage of all edges in both the directions. 
But it is still subject to the constraint that during each round, we can use each edge only once (in any direction that we intend to).
\begin{theorem}
Consider an undirected $(m,h)$-CCN with a $1$-node adversary. 
An outer bound for secrecy rate is $h-1$. 
Moreover, when\footnote{In the trivial case of $m=h$, secrecy rate $h-1$ is achievable using secure network coding \cite{secure_network_coding}.} $m\geq h+1$, there exists a scheme that achieves secrecy rate $(h-1)\frac{ m-h+1}{ m-h+2}$.
\end{theorem} 

For the above scheme, as $m\rightarrow \infty$, secrecy rate $\rightarrow h-1$. 
This shows asymptotic optimality of the scheme. 
Additionally, when $m\geq 2h-1$, the scheme achieves a secrecy rate strictly better than the outer bound $\frac{(h-1)^2}{h}$ for directed $(m,h)$-CCN.
Thus, feedback improves secrecy rates as we transition from directed CCNs to undirected CCNs.

\begin{IEEEproof} The outer bound proof is similar to (\ref{eq:SMT_undirected}) and follows from a mincut comprising of $h$ edges between a receiver and nodes in $\{B_1,B_2,\ldots,B_m\}$. 

\par We now show a simple scheme for an undirected $(m,h)$-CCN which achieves secrecy rate  $(h-1)\frac{ m-h+1}{ m-h+2}$. 
The scheme operates in two phases: uplink and downlink. 
It begins with a single round uplink phase where keys are collected at $S_1,S_2,\ldots,S_h$ as follows.
\begin{itemize}
\item Source $S$ sends keys $k_1^S,k_2^S,\ldots,k_{h-1}^S$ to $S_1,S_2,\ldots,S_{h-1}$ respectively and $k_h^S=-\sum_{i=1}^{h-1}k_i^{S}$ to $S_{h}$. This constitutes key set $\mathcal{K}^{S}$.
\item The receiver connected to $B_1,B_2,\ldots,B_h$ sends keys $k_1^R,k_2^R,\ldots,k_{h-1}^R$ to $S_1,S_2,\ldots,S_{h-1}$ respectively and $k_h^R=-\sum_{i=1}^{h-1}k_i^{R}$ to $S_{h}$. This constitutes key set $\mathcal{K}^{R}$.
\item For $i\in \{h+1,h+2,\ldots,m\} $, $A_i$ sends keys $k_1^{A_i},k_2^{A_i},\ldots,k_{h-1}^{A_i}$ to $S_1,S_2,\ldots,S_{h-1}$ respectively and $k_h^{A_i}=-\sum_{i=1}^{h-1}k_i^{A_i}$ to $S_{h}$. This constitutes key set $\mathcal{K}^{A_i}$.
\end{itemize}
The uplink phase is followed by a downlink phase comprising of $(m-h+1)$ downlink rounds. 
Each round of the downlink phase is similar to the second round of KSC scheme.
The key sets collected at $S_1,S_2,\ldots,S_h$ in the uplink phase are used as part of one-time pad and cancelled at non-trivial coding nodes (before they reach the receivers) in the following manner.  
\begin{enumerate}  
\item Key set $\mathcal{K}^{S}$ is cancelled at all non-trivial coding nodes in the first downlink round.
\item Key set $\mathcal{K}^{R}$ is cancelled at all non-trivial coding nodes in the second downlink round.
\item For the next $m-h-1$ downlink rounds, a key set from $\{ \mathcal{K}^{A_i} \}_{i\neq j}$ is cancelled at non-trivial coding node $A_j$.
\end{enumerate}
Each downlink round delivers $h-1$ symbols to all receivers. 
A single round uplink phase is followed by $(2+ m-h-1)$ downlink rounds and hence, the secrecy rate is $(h-1)\frac{ m-h+1}{ m-h+2}$.
\end{IEEEproof}

\subsection{Bidirected CCN}
We now consider a bidirected CCN, which we create by adding for every forward edge, one parallel edge (backward edge) of the opposite directionality. Note that this does not increase the mincut to the receivers. 
\begin{lemma}Consider a bidirected $(m,h)$-CCN with a 1-node adversary. When $m\geq h+1$, there exists a scheme which achieves secrecy rate $h-1$.
\end{lemma}

\begin{IEEEproof}
A single round scheme achieves secrecy rate $h-1$ as follows. The receiver connected to $B_1,B_2,\ldots,B_h$ first sends keys $k^R_1,k^R_2,\ldots,k^R_{h-1}$ to $S_1,S_2,\ldots,S_{h-1}$ and $k^R_h=-\sum_{i=1}^{h-1}k^R_i$ to $S_h$. These keys are then cancelled at non-trivial coding nodes (similar to the second round of KSC scheme) and $h-1$ symbols delivered to each receiver in the same round.
\end{IEEEproof}
Up to now we have focused on a $1$-node adversary. Interestingly, feedback using backward edges can help in the case of edge-adversaries as well. 
\begin{lemma}
For bidirected $(m,h)$-CCN, secrecy rate $h$ is achievable against a $1$-edge adversary (taps only one directed edge).
\end{lemma}
\begin{IEEEproof}
For every pair of nodes sharing an edge, a key can be sent using the parallel backward edge. This key can be used as a one-time pad to secure the network code on the forward edge. 
\end{IEEEproof}

\section*{Acknowledgment}
This work was supported in part by NSF award 1136174, MURI award AFOSR FA9550-09-064, ERC  Project NOWIRE ERC-2009-StG-240317 and a Ramanujan Fellowship from the DST (India).

\appendix \label{sec:appendix}
\subsection{Proof of Lemma~\ref{lemma:inner_tight_directed}}
We first give an intuitive outline of the achievability schemes followed by an illustrative example for directed $(4,3)$-CCN. The example is then extended to show schemes for the claimed cases.

\paragraph*{Intuitive outline} The scheme is based on the following observation. 
If we reduce the indegree of non-trivial coding nodes to $1$, a secure network code \cite{secure_network_coding} against $1$-edge adversary is sufficient to ensure secrecy.
But reducing the indegree of non-trivial coding nodes also reduces the mincut to some receivers, hence the secrecy rate.
Our approach is to have a multiple round routing strategy, using a different subset of edges in each round.
We still restrict the indegree of non-trivial coding nodes to $1$ in each round, but ensure that the mincut to each receiver averaged over multiple rounds is sufficient to achieve the desired secrecy rate. This has connection to the work in \cite{chekuri_average} on average throughput maximization (without any secrecy constraints) using tree packing strategies. The following example illustrates our approach for a directed $(4,3)$-CCN.

\begin{figure}[!th]
\begin{center}
\includegraphics[width=1.61in,height=2.1in]{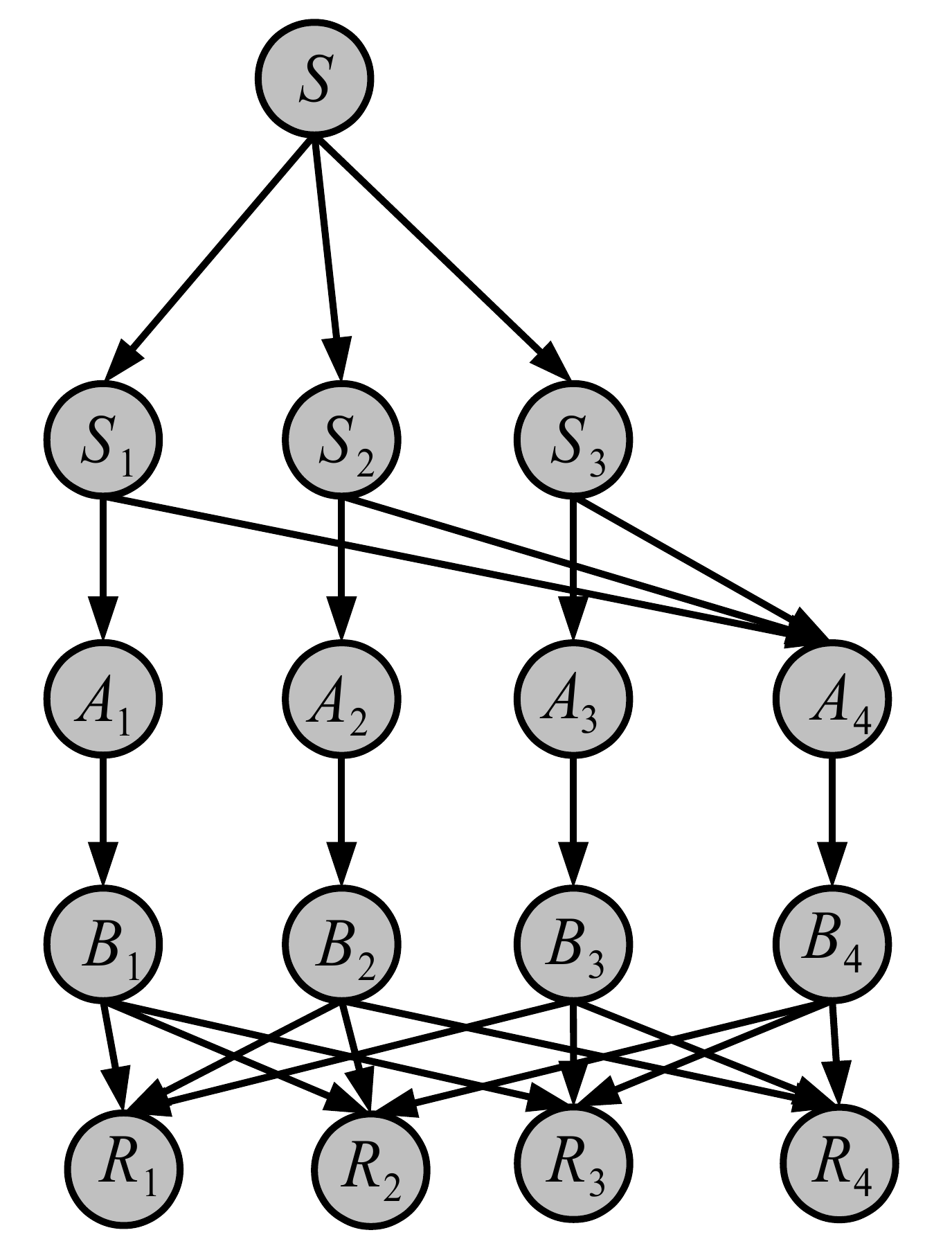}
\caption{Directed $(4,3)$-CCN.} %are accompanied by the cardinality of edges in the cut.}
\label{fig:routing_4_3}
\end{center}
\end{figure}

\begin{example} \label{ex:routing}
Consider a directed $(4,3)$-CCN with $4$ receivers as shown in Figure~\ref{fig:routing_4_3}. Let $\{c^1_1,c^1_2,c^2_1,c^2_2,c^3_1,c^3_2\}$ be the $6$ symbol codeword derived from a $4$ symbol message $\mathcal{W}$ using a rate $\frac{2}{3}$ erasure code. Let $\delta^1,\delta^2,\delta^3$ be keys generated by source $S$. We now describe a $3$-round scheme that achieves secrecy rate $\frac{4}{3}$ (optimal for directed $(4,3)$-CCN).
In the first round, $S$ sends $c^1_1+ \delta^1$, $c^1_1+c^1_2+\delta^1$ and $c^1_1-c^1_2+\delta^1$ to $S_1$,$S_2$ and $S_3$ respectively\footnote{These linear combinations are such that $c_2^1$ can be decoded from any two combinations, while $(c_1^1,c_2^1)$  can be decoded using all three combinations.}.
Nodes $S_1,S_2$ and $S_3$ send these values to $B_1,B_2$ and $B_3$ via $A_1,A_2$ and $A_3$. For $A_4$, $S_1 \rightarrow A_4$ is the only incoming edge used in this round and  $c^1_1+ \delta^1$ is sent from $S_1$ to $B_4$ via this edge. Each $B_i$ now sends the received values to the receivers connected to it. At the end of this round, receivers $R_1$  and $R_4$ decode $(c^1_1,c^1_2)$.
%-----------------------------------------------------------
The other two receivers decode only $c^1_2$.  In the second and third round, $\{c^1_1,c^1_2, \delta^1 \}$ are replaced with  $\{c^2_1,c^2_2, \delta^2\}$ and $\{c^3_1,c^3_2, \delta^3\}$ respectively with the following change in routing strategy for $A_4$. In the second round, $S_2 \rightarrow A_4$  is the only incoming edge used for $A_4$ and $c^2_1+c^2_2+\delta^2$ is sent along this edge. Hence, in the second round receivers $R_1$ and $R_3$ decode $(c^2_1,c^2_2)$  and the other two receivers decode $c^2_2$. In the third round, $S_3 \rightarrow A_4$ is the only incoming edge used for $A_4$ and $c^3_1-c^3_2+\delta^3$ is sent along this edge. In this round, receivers $R_1$ and $R_2$ decode $(c^3_1,c^3_2)$ and the other two receivers decode $c^3_2$. At the end of $3$ rounds, all receivers decode at least $4$ codeword symbols and hence decode the message. %In each round $i$, the presence of keys ($\delta^i$ in round $i$) ensures secrecy of all codeword symbols (and hence the message) at intermediate nodes.
%-----------------------------------------------------------
%---------------------------------------------------------------
\end{example}

\paragraph*{Extension to $(m\leq 6 ,h=3)$} In Example~\ref{ex:routing}, the values sent (over $3$ rounds) on edges $A_i \rightarrow B_i$ for $1\leq i \leq m$ can be listed in terms of $T^N_i$ (defined in proof of Theorem~\ref{thm:outer_bound_directed}, see Figure~\ref{fig:butterfly_proof})
\begin{eqnarray}
&&\left( \matrix{ T^3_1&T^3_2 &T^3_3&T^3_4 }\right) \nonumber \\
&= & \left( \matrix{c^1_1+\delta^1 & c^1_1+c^1_2+\delta^1 &   c^1_1-c^1_2+\delta^1  & c^1_1+ \delta^1 \cr  c^2_1+ \delta^2 & c^2_1+c^2_2+\delta^2 &  c^2_1 - c^2_2+\delta^2 &c^2_1+c^2_2+\delta^2  \cr  c^3_1+ \delta^3 & c^3_1+c^3_2+\delta^3& c^3_1-c^3_2+\delta^3& c^3_1-c^3_2+\delta^3}\right) \nonumber
\end{eqnarray}
Since the linear combinations used in each round are similar, we denote $c^i_1+\delta^i , c^i_1+c^i_2+\delta^i, c^i_1-c^i_2+\delta^i$ in round $i$ by \textit{routing symbols} $\underline{0}, \underline{1},\underline{2}$ respectively. In this notation,
\begin{eqnarray}
\left( \matrix{ T^3_1&T^3_2 &T^3_3&T^3_4 }\right) & \equiv&  \left( \matrix{\underline{0}&\underline{1}&\underline{2}&\underline{0}\cr \underline{0}&\underline{1}&\underline{2}&\underline{1}  \cr \underline{0}&\underline{1}&\underline{2}&\underline{2}}\right)  \nonumber\\
&=& R_{m=4,h=3,N=3}
%\left( \matrix{0&1&2&0&1 &2 \cr 0&1&2&1&2 &0  \cr 0&1&2&2&0 &1 }\right) 
\end{eqnarray} 
where $R_{m,h,N}$ is a \textit{routing matrix} listing the routing symbols sent on edges $A_i \rightarrow B_i$ as defined above. With this notation, we are now ready to compactly describe our optimal scheme for $(m=6,h=3)$. We simply extend $R_{m=4,h=3,N=3}$ to $R_{m=6,h=3,N=3}$ by adding two columns as shown below.
\begin{eqnarray}
R_{m=6,h=3,N=3}=\left( \matrix{\underline{0}&\underline{1}&\underline{2}&\underline{0}&\underline{1} &\underline{2} \cr\underline{0}&\underline{1}&\underline{2}&\underline{1} &\underline{2} &\underline{0}  \cr \underline{0}&\underline{1}&\underline{2}&\underline{2} &\underline{0} &\underline{1} }\right) 
\end{eqnarray}
Substituting back the values of $\underline{0}, \underline{1},\underline{2}$ for every round, one can easily check that every receiver (there are $6 \choose 3$ receivers) decodes at least $4$ codeword symbols over $3$ rounds and hence the secrecy rate is $\frac{4}{3}$ (optimal).

\paragraph*{Extension to $(m,h=2)$ based on Hadamard code} Let $m= 2N$ (if $m$ is odd, simply consider a directed $(m+1,h=2)$-CCN and proceed\footnote{The set of receivers in a directed $(m+1,h)$-CCN includes the receivers in a directed $(m,h)$-CCN; hence it is sufficient to show a scheme for (m+1,h).}). Let $\{c^1,c^2,\ldots,c^N\}$ be the $N$ symbol codeword derived from a $\frac{N}{2}$ symbol message using a rate $\frac{1}{2}$ erasure code. In addition, the source generates keys $\delta^1,\ldots,\delta^N$. We will now describe an $N$-round scheme which achieves secrecy rate $\frac{1}{2}$. In round $i$, we denote $c^i + \delta^i $, $\delta^i$ by routing symbols $\underline{0},\underline{1}$ respectively. In round $i$, the source sends  $\underline{0},\underline{1}$ to $S_1$,$S_2$ respectively and these get forwarded to $B_1$, $B_2$. This fixes the first two columns (corresponding to $T^N_1$ and $T^N_2$) of routing matrix $R_{m=2N,h=2,N}$ as the all $\underline{0}$ column vector and all $\underline{1}$ column vector respectively. The routing symbols sent on $A_3 \rightarrow B_3,\ldots,A_m\rightarrow B_m$ over $N$ rounds are derived from a Hadamard code as follows. Consider $2N$ Hadamard codewords (in terms of routing symbols $\underline{0},\underline{1}$) of length $N$ such that the all $\underline{0}$ and all $\underline{1}$ codewords are present in this collection (this can be easily done using Sylvester's construction). We assign these $2N$ codewords as the $2N$ columns of $R_{m=2N,h=2,N}$ such that the all $\underline{0}$ codeword is the first column and the all $\underline{1}$ codeword is the second column. The $N$-round schemes follows this routing matrix (the values sent on edges $A_i \rightarrow B_i$) and at the end of every round, $B_1,\ldots,B_{h+1}$ forward the received values to all the receivers connected to them. Since the Hamming distance between any two column vectors in $R_{m=2N,h=2,N}$ is at least\footnote{Property of Hadamard code.} $\frac{N}{2}$, each receiver receives both $\underline{0},\underline{1}$ (\emph{i.e.}, $c^i + \delta^i $, $\delta^i$ ) in at least $\frac{N}{2}$ rounds. Hence it can decode at least $\frac{N}{2}$ codeword symbols and thus the message.

\paragraph*{Extension to $(m=h+1,h)$} The scheme in Example~\ref{ex:routing} can be extended for the case $(h+1,h)$ as follows. Let $\{c^1_1,c^1_2\ldots,c^1_{h-1},c^2_1,\ldots,c^2_{h-1},\ldots,c^h_1,\ldots,c^h_{h-1} \}$ be the $(h-1)h$ symbol codeword derived from a $(h-1)^2$ symbol message using a rate $\frac{h-1}{h}$ erasure code. The source computes $\{x^i_1,\ldots,x^i_{h-1}\}$ from $\{c^i_1,\ldots c^i_{h-1}\}$ using an invertible linear transformation defined below.
\begin{eqnarray}
c^i_j = \sum_{l=1}^{h-1}x^i_l  - x^i_j 
\end{eqnarray}
In addition to the above steps, the source generates keys $\delta^1,\ldots,\delta^h$. We now describe an $h$-round scheme that achieves secrecy rate $\frac{(h-1)^2}{h}$.
In round $i$, for $1 \leq j \leq i-1$, $x^i_j + \delta^i$ is sent from $S$ to $S_{j}$ and for $i+1 \leq j \leq h$, $x^i_{j-1} + \delta^i$ is sent to $S_j$. For $j=i$, $\sum_{l=1}^{h-1} x^i_l + \delta^i$ is sent to $S_i$. These values are forwarded by $S_1,\ldots,S_h$ to $B_1,\ldots,B_h$ respectively via $A_1,\ldots,A_h$. Also, $S_i \rightarrow A_{h+1}$ is the only incoming edge used for $A_{h+1}$ in round $i$ and using this edge $\sum_{l=1}^{h-1} x^i_l + \delta^i$ is forwarded to $B_{h+1}$ via $A_{h+1}$. At the end of every round, $B_1,\ldots,B_{h+1}$ forward the received values to all the receivers connected to them. In round $i$, only two receivers can decode $h-1$ codeword symbols, i.e., $\{c^i_1,\ldots,c^i_{h-1}\}$ (they are the receivers connected to $\{B_1,\ldots,B_h\}$ and $\{B_1,\ldots,B_{h+1}\} -\{B_i\}$). The remaining $h-1$ receivers can decode only $h-2$ codeword symbols from $\{c^i_1,\ldots,c^i_{h-1}\}$. Hence over $h$ rounds, all receivers can decode at least $(h-1)(h-2) + (h-1) =(h-1)^2$ codeword symbols and achieve secrecy rate $\frac{(h-1)^2}{h}$.


\begin{thebibliography}{1}
\bibitem{maurer}U. M. Maurer, ``Secret key agreement by public discussion from
common information,'' IEEE Transactions on Information Theory 39(3), pp. 733-742, 1993.
\bibitem{secure_network_coding}N. Cai and R. W. Yeung, ``Secure network coding on a wiretap network,'' IEEE Transactions on Information Theory 57(1), pp. 424- 435, 2011.
\bibitem{koetter} R. Koetter and F. Kschischang, ``Coding for errors and erasures in
random network coding,'' IEEE Transactions on Information Theory 54(8), pp. 3579-3591, Aug. 2008.
\bibitem{netcod_12_ghid}Y. Buyukalp, G. Maatouk, V. Prabhakaran and C. Fragouli, ``Untrusting network coding,'' in Proc. IEEE Int. Symp. on Network Coding (NetCod), pp. 79-84, 2012.
\bibitem{kosut} O. Kosut, L. Tong, and D. Tse, ``Polytope codes against adversaries in networks,'' in Proc. IEEE Int. Symp. Inf. Theory (ISIT), pp. 2423-2427, 2010.
\bibitem{NCmonograph}C. Fragouli and E. Soljanin, ``Network Coding Fundamentals,'' Foundations and Trends in Networking, 2007.
\bibitem{vazirani}K. Jain, V. V. Vazirani and G. Yuval, ``On the capacity of multiple unicast sessions in undirected graphs,'' IEEE Transactions on Information Theory 52(6), pp. 2805-2809, 2006.
\bibitem{kamal_jain}K. Jain, ``Security based on network topology against the wiretapping attack,'' IEEE Wireless Communications, pp. 68-71, Feb 2004.
\bibitem{desmedt_revisited}Y. Wang and Y. Desmedt, ``Perfectly secure message transmission revisited,'' IEEE Transactions on Information Theory 54(6), pp. 2582-2595, 2008.
\bibitem{tracey_ho}T. Cui, T. Ho and J. Kliewer, ``On secure network coding with nonuniform or restricted wiretap sets,'' IEEE Transactions on Information Theory 59(1), pp. 166-176, 2013.
\bibitem{chekuri_average}C. Chekuri, C. Fragouli and E. Soljanin, ``On average throughput and alphabet size in network coding,'' IEEE Transactions on Information Theory 52(6), pp. 2410-2424, 2006.
%\bibitem{extended_version}S. Mishra, C. Fragouli, V. Prabhakaran and S. Diggavi, ``Using feedback for  secrecy over graphs,'' Extended version, Available: https://sites.google.com/site/shaunakmishracomm/isit13ext
\end{thebibliography}
\end{document}